\documentclass[12pt]{article}
\input epsf      
\usepackage{graphicx}% Include figure files

\usepackage{amssymb}
\usepackage{amsfonts}
\usepackage{amsthm}
\usepackage{amsfonts}

\newcommand{\half}{\mathbb{H}}

\newcommand{\compose}{\circ}

\newcommand{\ra}{\rightarrow}

\newcommand{\ba}[1]{\begin{array}{#1}}
\newcommand{\ea}{\end{array}}

\newcommand{\be}{\begin{equation}}
\newcommand{\ee}{\end{equation}}
\newcommand{\bea}{\begin{eqnarray}}
\newcommand{\eea}{\end{eqnarray}}
\newcommand{\beann}{\begin{eqnarray*}}
\newcommand{\eeann}{\end{eqnarray*}}

\newcommand{\reals}{{\mathbb R}}
\newcommand{\complex}{{\mathbb C}}

\newcommand{\blength}{b}
\newcommand{\nterms}{n}

\def\reff#1{(\ref{#1})}
\begin{document}

\title{Numerical computations for the Schramm-Loewner Evolution} 

\author{Tom Kennedy
\\Department of Mathematics
\\University of Arizona
\\Tucson, AZ 85721
\\ email: tgk@math.arizona.edu
}

\maketitle

\begin{abstract}

We review two numerical methods related to the Schramm-Loewner evolution
(SLE). The first simulates SLE itself. More generally, it finds
the curve in the half-plane that results from the Loewner equation 
for a given driving function. The second method can be thought of as the 
inverse problem. Given a simple curve in the half-plane it computes the 
driving function in the Loewner equation. This algorithm can be 
used to test if a given random family of curves in the half-plane 
is SLE by computing the driving process for the curves and testing if 
it is Brownian motion. More generally, this algorithm can be used
to compute the driving process for random curves that may not 
be SLE. Most of the material presented here has appeared before. 
Our goal is to give a pedagogic review, illustrate some of the 
practical issues that arise in these computations and discuss some 
open problems.

\end{abstract}

\bigskip

\newpage

\section{Introduction}

This review is about two types of numerical calculations
related to the Schramm-Loewner evolution (SLE). 
The first is to simulate SLE itself. More generally, one 
can consider simulating the random curves you obtain in the plane
when a random process is used for the driving function 
in the Loewner equation. 
The second type of simulation is to take a family of random curves
in the plane and compute the random driving process that 
generates them through the Loewner equation. This is related 
to SLE since one can test if a given family of random curves is SLE
by testing if the random driving process is Brownian motion. 
More generally, it is of interest to study the random driving 
process for random curves that may not be SLE. 
This review is meant to be pedagogic. Most of this material 
has appeared elsewhere. Our goal is to provide the reader 
with a ``how-to'' guide that will enable him or her to do 
state of the art simulations related to SLE. 

In the next section we give a heuristic and somewhat atypical
introduction to SLE that does not involve the Loewner equation. 
This is followed in section \ref{sect_LE} with a quick review 
of the Loewner equation and the usual definition of SLE. 
The ``discretization'' of SLE that is used in section 
\ref{sect_sle_intro} and discussed in detail in 
section \ref{sect_LE} was
studied extensively in \cite{bauer} for a particular approximation 
of the driving function (vertical slits). 
Reviews of SLE from the mathematics point of view include 
\cite{lawler_book,werner} and from the physics point of view include
\cite{bb_review,cardy_review,nienhuis_review}.

In section \ref{sect_drive_to_curve} we consider the numerical 
algorithm for finding a curve for a given driving function in 
the Loewner equation. Doing this with samples of Brownian 
motion for the driving function gives a simulation of SLE.

In section \ref{sect_curve_to_drive} we consider the numerical 
algorithm for finding the driving function for a given curve. 
One motivation for doing this is that it gives a way to test if a 
given model is SLE by testing if the driving process is Brownian motion.
Several works have considered models for which the connection with 
SLE is not clear, including domain walls in spin glasses 
\cite{ahhm,bdm} and turbulence \cite{bbcfA,bbcfB}.
Another motivation is to study the driving process for 
massive scaling limits of off-critical models \cite{bb_off, cfn, nw}

Both of the numerical algorithms we study are closely related to 
the zipper algorithm \cite{kuh,mr}. This is an algorithm 
for numerically finding the conformal map of a given 
simply connected domain onto a standard domain such as the 
unit disc. Much of the work described in this review grew out of 
conversations with Don Marshall and Stephen Rohde. 

\section{An introduction to SLE}
\label{sect_sle_intro}

In this section we will give a heuristic introduction to SLE.
The standard definition of SLE uses the Loewner equation 
from complex analysis. 
We will give a different definition of the process that does not 
use the Loewner equation. This view of SLE is well known, but is not 
typically discussed in reviews of SLE. 
The approach to SLE that we present is closely related 
to the numerical algorithms we will discuss. In the next section we will
see how this approach is related to the usual definition using 
the Loewner equation.

\begin{figure}[!h]
\includegraphics{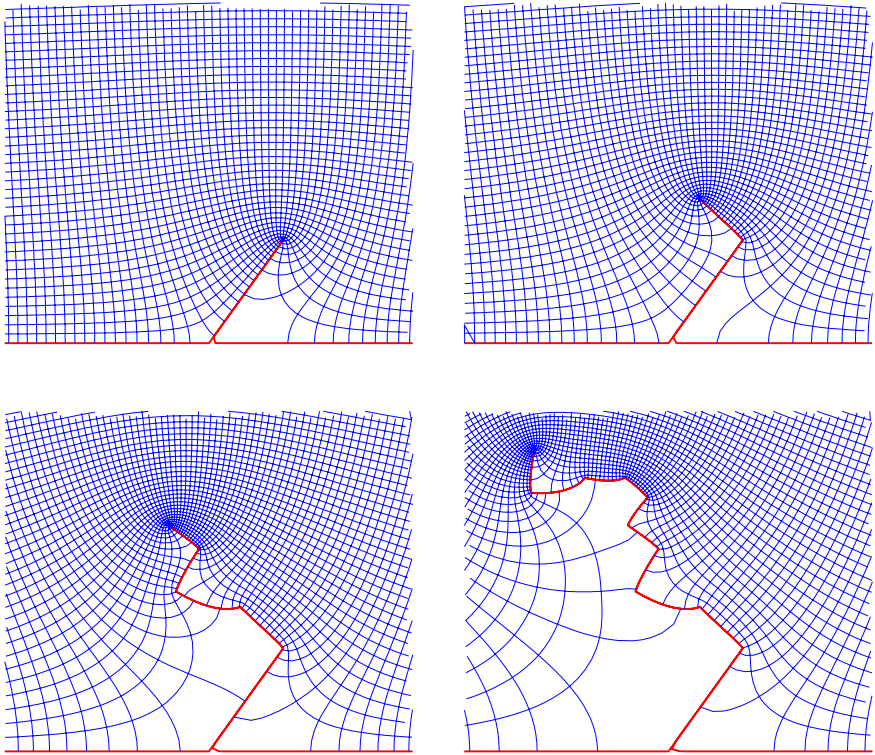}
\caption{\leftskip=25 pt \rightskip= 25 pt 
The figures illustrate the random composition of a sequence from the
maps $f_-$ and $f_+$. The numbers of maps in the compositions 
are $1,2,5$ and $10$.
}
\label{discrete}
\end{figure}

Let $\half$ denote the upper half of the complex plane, 
\bea
\half = \{ z : Im(z)>0\}
\eea
Fix an angle $\theta \in (0,\pi/2]$ and a length $\rho>0$.
Let $f_+(z)$ be the conformal map which takes $\half$ 
onto $\half \setminus \{ r e^{i\theta} : 0 < r \le \rho \}$, 
the upper half plane minus the line segment from $0$ to $\rho e^{i\theta}$.
This map is not unique. We make the choice unique by requiring 
\bea
f_+(\infty) &=& \infty \nonumber \\
f_+^\prime(\infty) &=& 1 \nonumber \\
f_+(0) &=& \rho e^{i \theta}
\label{f_norm}
\eea
The first two conditions mean that the Laurent series of $f_+$ about
$\infty$ is of the form
\beann
f_+(z)=z+ c_0 + {c_1 \over z} + {c_2 \over z^2} + \cdots 
\eeann
(For the reader familiar with the Loewner equation, we note that this 
is not the ``hydrodynamic'' normalization which would require that $c_0=0$
in the Laurent expansion instead of the 
third condition in \reff{f_norm}.)
The map $f_+$ is illustrated by the upper left picture in figure 
\ref{discrete}.
The grid shown is the image under the conformal map of the uniform 
rectangular grid in the upper half plane. 
Let $f_-(z)$ be the analogous conformal map for the segment from $0$ to 
$\rho e^{i(\pi-\theta)}$. (So the range of $f_-$ is the reflection of the range
of $f_+$ about the vertical axis.)

Consider composing two of these maps, e.g., $f_+ \compose f_-$. 
The effect of the second map in the composition will 
be to push the line segment created by the first map into the upper 
half plane and bend it somewhat. 
Because we have required that these maps send $0$ to the tip of 
the line segment, the lower endpoint of the
image of the first slit under the second map 
will be the tip of the second slit.  
In other words the image of $\half$ under the composition 
will be $\half$ with a curve removed. 
The map $f_+ \compose f_-$ is illustrated by the picture in the upper right
of figure \ref{discrete}. 

We can compose 
multiple copies of $f_-$ and $f_+$ and the resulting conformal map
will send the half plane onto the half plane minus a curve. 
We choose the maps randomly. 
Let $X_n$ be a sequence of independent, identically distributed random
variables with $X_n = \pm 1$ with probability $1/2$. 
For positive integers $n$ consider the conformal map
\bea
F_n = f_{X_1} \compose f_{X_2} \compose f_{X_3} \compose \cdots \compose 
f_{X_n} 
\label{fcomp}
\eea
(There is a slight abuse of notation here: $f_{\pm 1}$ means 
$f_{\pm}$.) The picture in the lower left of figure \ref{discrete} 
illustrates an example of $F_5$, and 
the picture in the lower right an example of $F_{10}$.

The conformal map $F_n$ will map $\half$ onto $\half \setminus \hat{\gamma}_n$
where $\hat{\gamma}_n$ is a curve in the upper half plane starting at $0$. 
Because of the order of the $X_i's$ in \reff{fcomp},
the curve $\hat{\gamma}_{n+1}$ will be an extension of the curve 
$\hat{\gamma}_n$. 
So we can let $n \ra \infty$ to get an infinite curve $\hat{\gamma}$. 
The SLE curve is then obtained by taking the scaling limit $\rho \ra 0$. 
The angle $\alpha$ is related to the usual parameter $\kappa$ for SLE. 
Using results that will appear latter, one can show that the relation is
\beann
\kappa = {4 (2 \alpha-1)^2 \over \alpha (1-\alpha)}
\eeann

When we use a piecewise smooth approximation to the driving function
in the Loewner equation, the curve $\hat{\gamma}$ will be simple
(non-intersecting).
It is a subtle question whether the curves one obtains in the limit
$\rho \ra 0$ are simple. For $\kappa \le 4$, SLE produces a simple curve 
\cite{rs}, and it is natural to conjecture that $\hat{\gamma}$ converges to 
this SLE curve. We do not prove this, and we are not aware of any proof
in the literature. For $\kappa>4$ the random set produced by SLE is 
not even a curve \cite{rs}. It is generated by a non-simple
curve, called the SLE trace, in the sense that the SLE set 
at time $t$ is the complement of the 
unbounded connected component of the half plane 
minus the curve up to time $t$. 
It is natural to conjecture that $\hat{\gamma}$ converges in distribution
to the SLE trace, but again we do not prove this and are not aware of 
any proof in the literature. Closely related questions are considered in 
\cite{bauer}.

\section{The Loewner equation}
\label{sect_LE}

We will now quickly review the Loewner equation from complex analysis and 
see how it is related to the definition of SLE that we gave in the 
previous section.
The Loewner equation provides a means for encoding curves in the 
upper half plane that do not intersect themselves by a real-valued 
function. In fact, it applies to more general growth processes in the half 
plane, but for the moment we restrict our attention to curves.
Let $\gamma(t)$ be a simple 
curve which lies in $\half$ for $0 < t < \infty$ and starts at the origin,
i.e., $\gamma(0)=0$.  
Let $\gamma[0,t]$ denote the image of $\gamma$ up to time $t$. Then 
$\half \setminus \gamma[0,t]$ is a simply connected domain. So there is 
a conformal map $g_t$ from this domain to $\half$. 
This map is not unique. We choose the map that satisfies
\bea
g_t(z) = z + {C(t)\over z} + O({1 \over |z|^2}), \qquad z \ra \infty
\label{hydro}
\eea
The coefficient $C(t)$ is called the half-plane capacity of $\gamma[0,t]$. 
It is known to be increasing in $t$, so
we can parametrize the curve so that $C(t)=2t$. 
Then $g_t$ satisfies Loewner's differential equation 
\be
{\partial g_t(z) \over \partial t} = {2 \over g_t(z) - U_t},
\qquad g_0(z)=z
\label{loewnereq}
\ee
for some real valued function $U_t$ on $[0,\infty)$. 
This statement is not obvious, and we refer the reader to 
\cite{lawler_book} for a proof. 
The function $U_t$ is often called the driving function.
We emphasize that while $g_t(z)$ is complex valued, the driving 
function $U_t$ is real-valued.

Note that $g_t$ goes in the opposite direction of the maps in the previous
section, i.e., it sends the half plane with a curve deleted onto
the half plane while the previous maps sent the half plane onto the 
half plane minus a curve. We should also note that $g_t$ is normalized 
differently since the constant term in \reff{hydro} vanishes. 
So $g_t(\gamma(t))$ is not the origin. In fact it is $U_t$. 
(To be precise, $g_t(\gamma(t))$ is not defined since $\gamma(t)$ 
is on the boundary of the domain of $g_t$. Its image under $g_t$ 
must be defined by a limiting process.)

If our simple curve in the half plane is random, 
then the driving function $U_t$ is a stochastic process.
Schramm's wonderful discovery was 
that if the scaling limit of a two-dimensional 
model is conformally invariant and satisfies a property usually
called the domain Markov property, 
then this stochastic driving process must be a Brownian motion 
with mean zero \cite{schramm}. 
The only thing that is not determined is the variance. 
Schramm named this process stochastic Loewner evolution or SLE; it is now
often referred to as Schramm-Loewner evolution. 

The solution to \reff{loewnereq} need not exist for all times $t$ since
the denominator can go to zero. 
We let
$K_t$ be the set of points $z$ in $\half$  for which the 
solution to this equation no longer exists at time $t$. 
If we start with a simple curve and define $g_t$ as we did above, then
$K_t$ will be $\gamma[0,t]$. But if we start with a continuous 
driving function $U_t$ and solve the Loewner equation, $K_t$ will 
only be a curve for sufficiently nice $U_t$. (Just what sufficiently 
nice means is a subtle question \cite{mrc}.)
For other $U_t$, $K_t$ can be a more complicated growing set. 
In particular, when $U_t$ is a Brownian motion, $K_t$ may not be a curve. 
In our simulations, even in the cases where $U_t$ is not sufficiently nice,
our approximation to $U_t$ will be nice enough that it produces a curve. 
So in the following we will always take $K_t$ to be a curve, but the 
reader should keep in mind that in some cases this curve is 
approximating a more complicated set.

Let $t,s>0$. The map $g_{t+s}$ maps $\half \setminus \gamma[0,t+s]$ 
onto $\half$. 
We can do this in two steps. We first apply the map $g_s$. 
This maps $\half \setminus \gamma[0,s]$ onto $\half$, and it maps  
$\half \setminus \gamma[0,t+s]$ onto 
$\half \setminus g_s(\gamma[s,t+s])$. 
Let $\bar{g}_t$  be the conformal map that maps 
$\half \setminus g_s(\gamma[s,t+s])$ 
onto $\half$ with the usual hydrodynamic normalization. 
Then $\bar{g}_t \compose g_s$ will map $\half \setminus \gamma[0,t+s]$ onto 
$\half$ and satisfy \reff{hydro}. There is only one such conformal map, so  
\be
g_{s+t}=\bar{g}_t \compose g_s, \quad i.e., \quad 
\bar{g}_t = g_{s+t} \compose g_s^{-1}
\ee
It we think of $s$ as being fixed and $t$ as the time variable,
then the function $\bar{g}_t$  is also a solution of the Loewner
equation 
\be
{d \over dt} \bar{g}_t(z)
={d \over dt} g_{s+t} \compose g_s^{-1}(z) 
= {2 \over g_{s+t} \compose g_s^{-1}(z) -U_{s+t}}
= {2 \over \bar{g}_{t}(z)-U_{s+t}}
\ee
and satisfies $\bar{g}_0(z)=z$. Thus  $\bar{g}_t(z)$ is obtained by 
solving the Loewner equation with driving function $\bar{U}_t=U_{s+t}$.
This driving function starts at $U_s$, and so the curve 
associated with $\bar{g}_t$ starts at $U_s$.

We now introduce a partition of the time interval $[0,\infty)$: 
$0=t_0 < t_1 < t_2 < \cdots t_n < \cdots$, and 
define
\be
\bar{g}_k = g_{t_k} \compose g_{t_{k-1}}^{-1}
\ee
So 
\be 
g_{t_k} = \bar{g}_k \compose \bar{g}_{k-1} \compose \bar{g}_{k-2} \compose 
\cdots \compose
\bar{g}_2 \compose \bar{g}_1 
\label{composition}
\ee
By the remarks above, $\bar{g}_k$ is obtained by solving the Loewner equation
with driving function $U_{t_{k-1} +t}$ for $t=0$ to $t=\Delta_k$,
where $\Delta_k = t_k-t_{k-1}$. 
The image of $\half$ under $\bar{g}_k$ is 
$\half$ minus a ``cut'' starting at $U_{t_{k-1}}$.
So if we shift it by defining 
\bea 
g_k(z) = \bar{g}_k(z+U_{t_{k-1}}) - U_{t_{k-1}},
\eea 
then $g_k$ is obtained by solving the 
Loewner equation with driving function $U_{t_{k-1}+t} - U_{t_{k-1}}$
for $t=0$ to $t=\Delta_k$. This driving function starts at 
$0$ and ends at $\delta_k$ where $\delta_k=U_{t_k}-U_{t_{k-1}}$.
So this conformal map takes 
$\half$ minus a cut starting at the origin onto $\half$.
The inverse of this map, 
\bea 
g_k^{-1}(z) = \bar{g}_k^{-1}(z+U_{t_{k-1}}) - U_{t_{k-1}},
\eea 
takes $\half$ and introduces a cut which begins at the origin.

There are two general types of simulations we would like to do. 
Given a driving function we want to find the curve it generates. 
And given a curve we want to find the corresponding driving function. 
For both problems the key idea is the same. 
We approximate the driving function on the interval $[t_{k-1},t_k]$
by a function for which the Loewner equation may be explicitly solved. 
So the maps $\bar{g}_k$ and $g_k$ can be found explicitly. 
Eq. \reff{composition}
can then be used to approximate $g_t$. We will consider two 
explicit solutions of the Loewner equation which we will 
refer to as ``tilted slits'' and ``vertical slits.''

For tilted slits, let $x_l, x_r >0$ and $0<\alpha< 1$. Then define 
\beann
f(z)= (z+x_l)^{1-\alpha} (z-x_r)^{\alpha}, \qquad 
\eeann
Then $f$ 
maps $\half$ to $\half \setminus \Gamma$ where $\Gamma$ is a 
line segment from $0$ to a point $\rho e^{i\alpha \pi}$. 
The length $\rho$ can be expressed in terms of $x_l,x_r$ and $\alpha$. 
This map sends $[-x_l,x_r]$ onto $\Gamma$. 
Unfortunately, its inverse cannot be explicitly computed. 
For the inverse to satisfy the normalization 
\reff{hydro}, we must have 
\bea
(1-\alpha) x_l= \alpha x_r
\label{xcond}
\eea
Straightforward calculation shows if we let 
\beann
f_t(z)=\left(z+ 2 \sqrt{t} \sqrt{\alpha \over 1-\alpha} \right)^{1-\alpha} 
\left(z- 2 \sqrt{t} \sqrt{1-\alpha \over \alpha} \right)^\alpha
\eeann
then it produces a slit with capacity $2t$. 
We know that $g_t=f_t^{-1}$ must
satisfy the Loewner equation \reff{loewnereq}
for some driving function $U_t$. More calculation shows that the 
driving function is 
\be
U_t = c_\alpha \sqrt{t}, \qquad
c_\alpha=2 {1-2 \alpha \over \sqrt{\alpha (1-\alpha)}}
\ee
The change in the driving function over the time interval $[0,\Delta]$ is 
\be
\delta = c_\alpha \sqrt{\Delta}
\label{alphaeq} 
\ee
The original map $\phi$ had three real degrees of freedom, 
$\alpha,x_l,x_r$. The condition  \reff{xcond} reduces this to 
two real degrees of freedom, $\alpha$ and $t$. 
So if we are given $\delta$ and $\Delta$ or given $\rho$ and $\alpha$, 
then the map is completely determined. 

Vertical slits correspond to an even simpler solution of 
the Loewner equation. Let 
\beann
g_t(z)=\sqrt{(z-\delta)^2+4t}+\delta
\eeann
Then it is easy to check that  $g_t$ satisfies Loewner's equation with 
a constant driving function, $U_t=\delta$.
Since the driving function does not start at $0$, the curve will 
not start at the origin. 
The curve is just a vertical slit from $\delta$ to $\delta + 2i \sqrt{t}$. 
Using vertical slits means that we approximate the driving function by a 
discontinuous piecewise constant function.
This will produce a $K_t$ which is not a curve.

Our numerical studies only use tilted slits and vertical slits 
for the explicit solutions for the Loewner equation. 
Another possibility is to use a linear driving function.
If we let $h_t = g_t - U_t$, then the differential equation for $h_t$ 
can be solved by separation of variables. The solution is not 
completely explicit - it must be expressed in terms of a function that is
defined implicitly by a transcendental equation.

\section{From the driving function to the curve}
\label{sect_drive_to_curve}

Our primary motivation is to simulate SLE, i.e., to compute the 
curve when the driving function is Brownian motion.
But our discussion is more general, and the following algorithm
can be used to calculate the curve corresponding to any driving function 
$U_t$. 

There are a variety of conformal maps that occur in this paper, and 
we have denoted them by letters that indicate what they do. 
Maps denoted with $g$ are solutions of the Loewner equation
with a driving function that starts at $0$. 
So they map the half plane minus a curve starting at the origin onto the 
half plane, sending the tip of the curve to the final value of the 
driving function. We use $\bar{g}$ for solutions to the Loewner equation
when the driving function does not start at $0$. In this case the curve
starts at the initial value of the driving function and the map 
still sends the tip to the final value of the driving function.
If we follow a map $g$ by a real translation that takes the final value
of the driving function to $0$, we get a map that takes 
the half plane minus a curve onto the half plane and sends 
the tip to the origin. We denote such maps by $h$. (Note that such 
maps do not satisfy the Loewner equation.) Finally, we use
$f$ to denote maps that are inverses of maps $h$. So they take 
the half plane onto the half plane minus a curve and sends 
the origin to the tip. 

Let $0=t_0 < t_1 < t_2 < \cdots < t_n$ be a partition of the time
interval $[0,t]$. 
The SLE curve is given by $\gamma(t)=g_t^{-1}(U_t)$.
Let $z_k=g_{t_k}^{-1}(U_{t_k})$.
We will only consider the points $z_k$ on this curve which correspond to 
times $t=t_k$. One could consider other points on the curve, 
but the distance between consecutive $z_k$ is already of the order of the 
error in our approximation, so there is no reason to consider more points. 
By \reff{composition} the points $z_k$ are given by 
\be 
z_k= \bar{g}^{-1}_1 \compose \bar{g}^{-1}_2 \compose \cdots \cdots 
\bar{g}^{-1}_{k-1} \compose \bar{g}^{-1}_k (U_{t_k}) 
\ee
Recall that if we solve the Loewner equation with driving function 
$U_{t_{k-1}+t} - U_{t_{k-1}}$ for $t=0$ to $t=\Delta_k$,
the result is $g_k(z)$ where 
\be
g_k(z)=\bar{g}_k(z+U_{t_{k-1}}) - U_{t_{k-1}}
\ee
Define 
\be
h_k(z)=g_k(z)- \delta_k = \bar{g}(z+U_{t_{k-1}})-U_{t_k}
\ee
where $\delta_k=U_{t_k}-U_{t_{k-1}}$. 
Then
\be
h_k \compose h_{k-1} \compose \cdots \cdots \compose h_1(z_k) =
\bar{g}_k \compose \bar{g}_{k-1} \compose \cdots \cdots \compose 
\bar{g}_1(z_k) - U_{t_k} =0. 
\ee
Let 
\be
f_k = h_k^{-1}
\ee
So 
\be
z_k = f_1 \compose f_2 \compose \cdots \compose f_k(0)
\label{dsle}
\ee

As noted before, $g_k$ maps $\half$ minus a small curve 
onto $\half$.
The driving function ends at $\delta_k$, so $g_k$
sends the tip of the curve to $\delta_k$. 
It follows that $h_k(z)=g_k(z)-\delta_k$ maps 
$\half$ minus the small curve onto $\half$ and sends the tip to the 
origin.
So $f_k=h_k^{-1}$ maps $\half$ onto $\half$ minus the small curve 
and sends the origin to the tip of the curve. 
Thus the functions $f_k$ are analogous to the functions $f_\pm$ from 
section \ref{sect_sle_intro} in that they all introduce a small cut into
the upper plane and send the origin to the tip of the cut. Note the 
similarity of \reff{dsle} to \reff{fcomp}.

As discussed before, we define $U_t$ on each time interval
$t_{k-1} \le t \le t_k$ so that $g_k(z)$ may be explicitly computed. 
There are two constraints
on $g_k$. The curve must have capacity $2 \Delta_k$ and 
$g_k$ must map the tip of the curve to $\delta_k$. 
Any simple curve satisfying these two constraints and starting at the 
origin will correspond to a solution of the Loewner equation for some 
driving function which goes from $0$ to $\delta_k$ over the time 
interval $[0,\Delta_k]$. So our approximation can be thought of 
as replacing the driving function by a new driving function that 
agrees with the original one at the times $t_k$ but differs 
in between those times. 

Different choices of how we define $U_t$ on each time interval 
give us different discretizations. 
As we will see, this choice will not have a significant effect. 
Of much greater importance is how we choose the 
$\Delta_k$ and $\delta_k$. 

\begin{figure}[!h]
\includegraphics{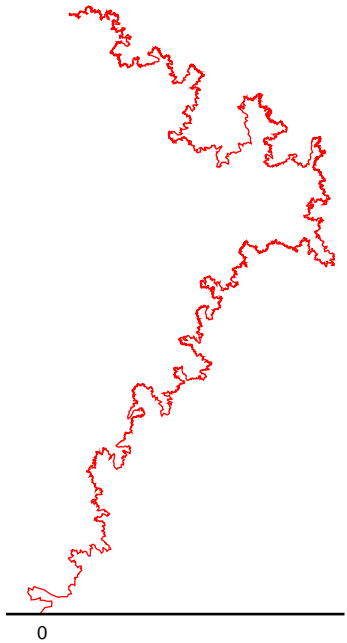}
\caption{\leftskip=25 pt \rightskip= 25 pt 
SLE with $\kappa=8/3$ with fixed $\Delta t$. There are $10,000$ points. 
}
\label{sle_fixed_2.66}
\end{figure}

If we want to simulate SLE, the $\delta_k$ 
should be chosen so that the stochastic process $U_t$
will converge to $\sqrt{\kappa}$ times Brownian motion as 
$N \rightarrow \infty$. 
One choice is take the $\delta_k$ to be independent normal random 
variables with mean zero and variance $\kappa \Delta_k$. If we do this, 
then $U_t$ and $\sqrt{\kappa} B_t$ will have the same distributions if 
we only consider the times $t_k$. 
Another possibility is to approximate the Brownian motion by a 
simple random walk. This is done by using a uniform partition of 
the time interval and 
taking the $\delta_k$ to be independent random variables with 
$\delta_k = \pm \sqrt{\kappa \Delta_k}$ where the choices of $+$ and $-$ both 
have probability $1/2$. This is what we were doing in section 
\ref{sect_sle_intro}.

\begin{figure}[!h]
\includegraphics{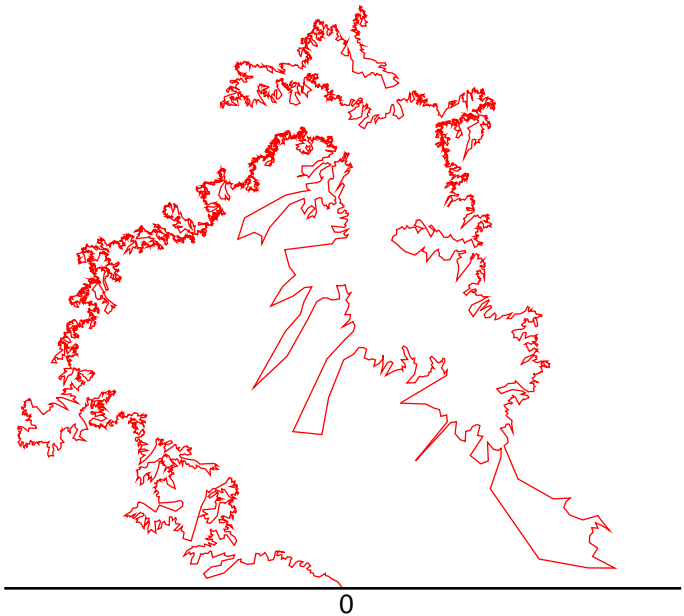}
\caption{\leftskip=25 pt \rightskip= 25 pt 
SLE with $\kappa=6$ with fixed $\Delta t$. There are $10,000$ points. 
}
\label{sle_fixed_6}
\end{figure}

\begin{figure}[!h]
\includegraphics{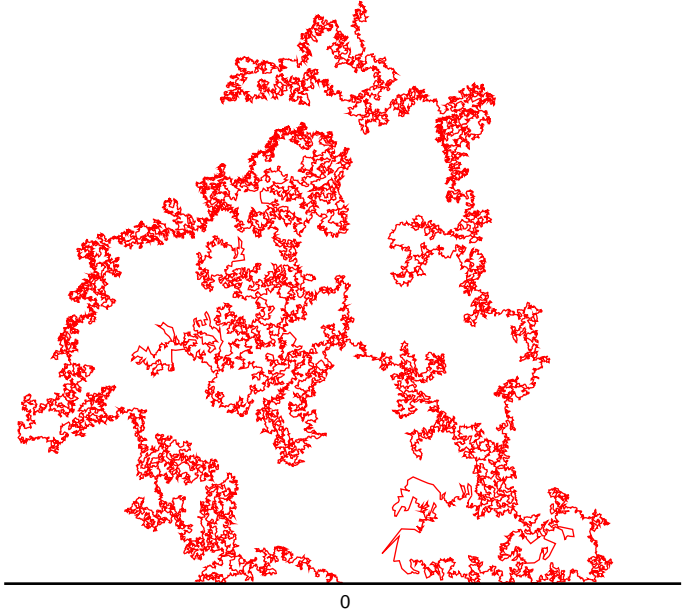}
\caption{\leftskip=25 pt \rightskip= 25 pt 
SLE with $\kappa=6$ with adaptive $\Delta t$. There are $35,000$ points. 
}
\label{sle_6}
\end{figure}

The simplest choice for $\Delta_k$ is to use a uniform partition of the 
time interval. For values of $\kappa$ which are not 
too large this works reasonably well. Figure \ref{sle_fixed_2.66} 
shows a simulation using $\kappa=8/3$ with $10,000$ equally spaced
time intervals. 
However, for larger values of $\kappa$, uniform $\Delta_k$ are 
a disaster. Figure \ref{sle_fixed_6} shows a simulation with $\kappa=6$ and 
$10,000$ equally spaced time intervals. Clearly something has gone wrong.
To see just how badly wrong things have gone the reader should 
compare this figure with figure  \ref{sle_6} which uses the same 
sample of Brownian motion. 

To understand the effect seen in figure \ref{sle_fixed_6} we 
give an equivalent definition of the 
half plane capacity $C$ of a set $A$. We originally defined it by
\beann
g(z) = z + { C \over z} + O({1 \over z^2})
\eeann
where $g$ maps $\half \setminus A$ onto $\half$.
A more intuitive definition is 
\beann
C = \lim_{y \ra \infty} y \, E^{iy}[Im(B_\tau)]
\eeann
where $B_t$ is two-dimensional Brownian motion started at $iy$.
The stopping time $\tau$ is the first time the 
Brownian motion  hits $A$ or $\reals$. 
From the point of view of this two-dimensional 
Brownian motion, parts of the curve
can be well hidden by earlier parts of the curve and so have 
very little capacity. So what looks like a ``long'' section of 
the curve has very little capacity and so gets very few 
points approximating it. 

To do better we will use non-uniform $\Delta_k$. 
In fact they will depend on the sample of the Brownian motion and 
so we refer to this method as ``adaptive $\Delta_k$.'' 
(I learned this idea from Stephen Rohde \cite{rohde}.)
Fix a spatial scale $\epsilon>0$. 
We start with a uniform partition of the time and compute the points $z_k$
along the curve. Then we look for points 
$z_k$ such that $|z_k-z_{k-1}| \ge \epsilon$.
For these time intervals $[t_{k-1},t_k]$, divide the interval 
into two equal intervals. 
We then sample the Brownian motion at the midpoint of $[t_{k-1},t_k]$ using 
a Brownian bridge. (This just means that to choose the value of the 
driving function at the midpoint of $[t_{k-1},t_k]$ we use a Brownian 
motion conditioned on the values we already have for it at $t_{k-1}$ 
and $t_k$.)
Then we recompute all the $z_k$. (There will of course be more
of them than before.) Note that we must recompute all the points since
even at times which appeared in the time partition before, the 
corresponding point on the curve will change.
We repeat this until we have $|z_k - z_{k-1}| \le \epsilon$ for all $k$.

\begin{figure}[tbh]
\includegraphics{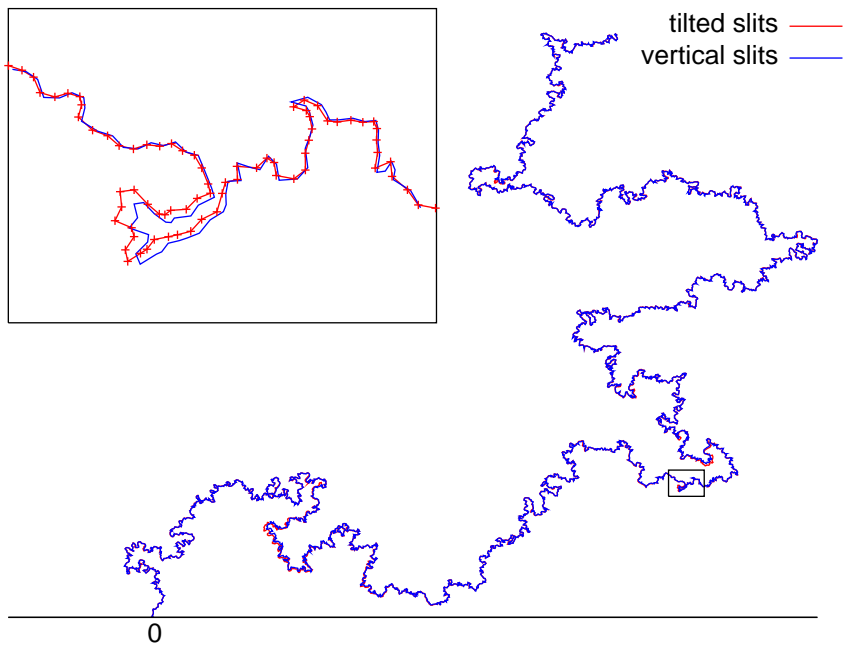}
\caption{\leftskip=25 pt \rightskip= 25 pt 
A comparison of the curves obtained using tilted slit 
maps and vertical slit maps with $\kappa=8/3$.}
\label{map_choice_2.66}
\end{figure}

Our approximation can be thought of as approximating the driving 
function by a concatenation of driving functions on short time intervals 
for which the Loewner equation is exactly solvable. It is important to 
consider the effect of the choice of which exactly solvable 
driving functions we use. 
To do this we compare the curves we get using tilted 
slits for the elementary maps with the curves we get using vertical 
slits. We carry out the adaptive simulation just described 
using tilted slits. We then use the 
same $\Delta_k$ and $\delta_k$, i.e., the same partition of the time 
interval and the same sample of Brownian motion, but with 
vertical slits. 
For $\kappa=8/3$, figure \ref{map_choice_2.66}
shows the tilted slits curve vs. the vertical slits curve. 
The vertical slits do not produce a curve. What we have plotted is 
the following. We compute the points $z_k$ and then just
connect them with a straight line.
In figure \ref{map_choice_2.66} it is almost impossible to 
distinguish the two curves. 
An enlargement of part of the curves is shown in the inset. 
Even in the enlargement the difference is quite small.
The curves have a relatively small number of points (about 6,000), and 
in the enlargement we have plotted the points for the tilted slit 
curve. The difference between the two curves is on the order of the 
distance between these points.

\begin{figure}[tbh]
\includegraphics{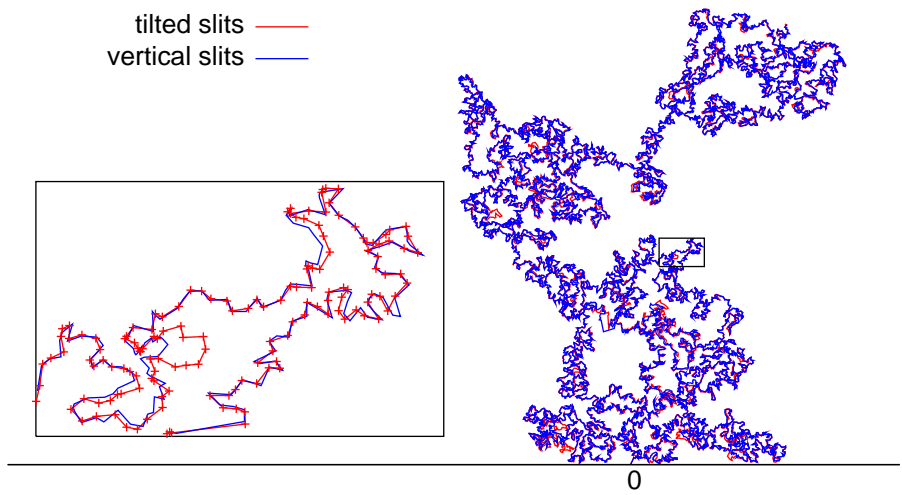}
\caption{\leftskip=25 pt \rightskip= 25 pt 
A comparison of the curves obtained using tilted slit 
maps and vertical slit maps with $\kappa=6$.}
\label{map_choice_6}
\end{figure}

Figure \ref{map_choice_6} shows 
the same thing with $\kappa=6$. In the enlargement one can see 
deviations between the two curves, but the size of the deviations is 
again on the same scale as the distance between adjacent points 
on the curve. 

It is interesting to note that there is what one might call a
stability to the approximation we are using. 
The difference between the two curves in figures 
\ref{map_choice_2.66} and \ref{map_choice_6} fluctuates with 
time, but it does not grow with time. In other words, the errors 
from approximating the true driving function over the short time
intervals do not appear to accumulate.

\clearpage

\section{From the curve to the driving function }
\label{sect_curve_to_drive}

We now consider what one might call the inverse problem. 
Given a simple curve $\gamma$, we want to compute the 
corresponding driving function. 

Let $\gamma(s)$ be a parametrized simple curve in $\half$. 
In almost all applications, the parametrization of the curve 
is not the parametrization by capacity. 
Let $g_s$ be the conformal map which takes $\half \setminus \gamma[0,s]$ 
onto $\half$, normalized so that for large $z$ 
\be
g_s(z) = z + { C(s) \over z} + O({1 \over z^2}),
\label{laurent_norm}
\ee
The coefficient $C(s)$ is the half-plane capacity of $\gamma[0,s]$. 
The value of the driving function at $t=C(s)/2$ is $U_t=g_s(\gamma(s))$.
Thus computing the driving function essentially reduces to 
computing this uniformizing conformal map. 

Let $z_0,z_1,\cdots,z_n$ be points along the curve $\gamma$ with $z_0=0$. 
In many applications these are lattice sites. 
We will find a sequence of conformal maps 
$h_i$, $i=1,2,\cdots,n$ such that 
$h_k \compose h_{k-1} \compose \cdots \compose h_1(z_k)=0$.
Then $h_k \compose h_{k-1} \compose \cdots \compose h_1$
sends $\half \setminus \hat{\gamma}$ to $\half$ where 
$\hat{\gamma}$ is some curve that passes through $z_0,z_1, \cdots z_k$
and so approximates $\gamma$.
Suppose that the conformal maps $h_1,h_2,\cdots,h_k$  have been 
defined with these properties.
Let 
\be
w_{k+1}=h_k \compose h_{k-1} \compose \cdots \compose h_1(z_{k+1})
\label{compose}
\ee
Then $w_{k+1}$ is close to the origin.
We define $h_{k+1}$ to be a conformal map that
sends $\half \setminus \gamma_{k+1}$ to $\half$
where $\gamma_{k+1}$ is a short simple curve from $0$ to $w_{k+1}$. 
We also require that $h_{k+1}$ sends $w_{k+1}$ to the origin. 
As before we choose the curve $\gamma_{k+1}$ so that $h_{k+1}$
is explicitly known; possible choices include
``tilted slits'' and ``vertical slits.'' 
Note that for both of these maps there were two real degrees of freedom. 
They will be determined by the condition that $h_{k+1}(w_{k+1})=0$.

Let $2 \Delta_i$ be the capacity of the map $h_i$, and $\delta_i$ 
the final value of the driving function for $h_i$. So 
\be
h_i(z) = z - \delta_i + { 2 \Delta_i \over z} + O({1 \over z^2})
\ee
Then 
\be
h_k \compose h_{k-1} \compose \cdots \compose h_1(z) =
z  - U_t + { 2 t \over z} + O({1 \over z^2})
\ee
where 
\be
t= \sum_{i=1}^k \Delta_i 
\ee
\be
U_t= \sum_{i=1}^k \delta_i 
\ee
Thus the driving function of the curve is obtained by concatenating 
the driving functions of the elementary conformal maps $h_i$. 

\section{Faster algorithms}
\label{sect_faster}

In this section we show how to speed up both the algorithm 
for computing the curve $\gamma$ given the driving function $U_t$ and 
the algorithm for computing the driving function $U_t$ given a curve
$\gamma$. We start with the first algorithm. One of the main motivations
is a fast algorithm for simulating SLE, but our fast algorithm 
is applicable to other driving functions as well. 

Recall that points on the approximation to the SLE trace or more 
generally the curve $\gamma$ are given by eq. \reff{dsle} which says 
\be
z_k= f_1 \compose f_2 \compose \cdots \compose f_k(0)
\label{composeb}
\ee
The number of operations needed to compute a single $z_k$ is
proportional to $k$. So to compute all the points $z_k$ with $k=1,2,\cdots N$
requires a time $O(N^2)$. 
The computation of $z_k$ does not depend on any of the other $z_j$. 
Depending on what we want to compute, we may only need to compute 
a subset of the points $z_k$.
(For example, if we are only interested in 
$z_N=\gamma(t_N)$, the time required is $O(N)$ not $O(N^2)$.) 
For a typical point $z_k$, the time to compute it is $O(N)$ for 
the above algorithm. Our goal is to develop an algorithm for 
which this time is $O(N^p)$ with $p<1$. 

Our algorithm groups the functions in \reff{composeb} into
blocks. We denote the number of functions in a block by $\blength$. 
Let
\be
F_j = f_{(j-1)b+1} \compose f_{(j-1)b+2} \compose \cdots \compose f_{jb}
\label{blockdefb}
\ee
If we write $k$ as $k=mb+l$ with $0 \le l < b$, then we have 
\be
z_k = F_1 \compose F_2 \compose \cdots \compose F_m \compose 
f_{mb+1} \compose f_{mb+2} \compose \cdots \compose f_{mb+l}(0) 
\label{blockcomposeb}
\ee
The number of compositions in \reff{blockcomposeb} is smaller than the 
number in \reff{composeb} by roughly a factor of $b$ if $b$ is smaller 
than $m$, i.e., if $k$ is bigger than $b^2$. 

 Unfortunately, 
even though the $f_i$ are explicit and relatively simple, 
the $F_j$ cannot be explicitly 
computed. Our strategy is to approximate the $f_i$ by functions 
whose compositions can be explicitly computed to give an explicit
approximation to $F_j$. For large $z$, $f_i(z)$ is given
by its Laurent series about $\infty$. One could approximate $f_i$ 
by truncating this Laurent series. 
Our approximation is of this nature, but slightly different.

Let $\gamma : [0,t] \rightarrow \half$ be a simple curve in the upper half
plane with $\gamma(0)=0$. 
Let $f(z)$ be the conformal map from $\half$ 
onto $\half \setminus \gamma[0,t]$.
We assume that $f$ is normalized is the same way as our $f_i$, i.e.,
$f(\infty)=\infty$, $f^\prime(\infty)=1$ and 
$f(0)=\gamma(t)$.
Let $a,b>0$ be such that $[-a,b]$ is mapped onto the slit $\gamma[0,t]$. 
Then $f$ is real valued on $(-\infty,-a] \cup [b,\infty)$, and so  
$f$ has an analytic continuation to $\complex \setminus [-a,b]$
by the Schwartz reflection principle. We denote this extension by just $f$. 

Let $R=\max \{ a,b\}$, so $f$ is analytic on $\{z: |z|>R\}$ and maps
$\infty$ to itself. Thus $f(1/z)$ is analytic on $\{z: 0<|z|<1/R\}$.
Since our assumptions on $f$ imply it has a simple pole at the origin with 
residue 1, we have 
\be
f(1/z)= 1/z + \sum_{k=0}^\infty \, c_k \, z^{k}
\ee
This gives the Laurent series of $f$ about $\infty$. 
\be
f(z)= z + \sum_{k=0}^\infty \, c_k \, z^{-k}
\ee
This Laurent series is a natural 
approximation to use for $f$ when $z$ is large. 
However, we will use a different but closely related representation.

Define $\hat{f}(z)=1/f(1/z)$. Since $f(z)$ does not vanish on 
$\{ |z|>R\}$, $\hat{f}(z)$ is analytic in $\{ z : |z| < 1/R \}$.
Our assumptions on $f$ imply that $\hat{f}(0)=0$ and $\hat{f}^\prime(0)=1$.
So $\hat{f}$ has a power series 
\be
\hat{f}(z) = z + \sum_{j=2}^\infty \, a_j z^j
\label{hpsb}
\ee
The radius of convergence of this power series is easily shown to be $1/R$.
Note that the coefficients of the power series of $\hat{f}$ are the
coefficients of the Laurent series of $1/f$. 

The primary advantage of our power series over the 
Laurent series is its behavior with respect to composition. 
\be
(f_1 \compose f_2) \, \hat{} \,(z) = {1 \over f_1((f_2(1/z))}
= { 1 \over f_1(1/\hat{f_2}(z))} = 
\hat{f_1}(\hat{f_2}(z))
\ee
Thus 
\be
(f_1 \compose f_2) \, \hat{} \, = \hat{f_1} \compose \hat{f_2}
\label{composepropb}
\ee
Our approximation for $f(z)$ is to approximate $\hat{f}(z)$ 
by the truncation of its power series at order $\nterms$. So 
\be
f(z) = {1 \over \hat{f}(1/z)}
\approx \left[ \sum_{j=0}^n \, a_j z^{-j} \right]^{-1}
\label{hat_approx}
\ee

For each $f_i$ we compute the power series of $\hat{f_i}$ to order $\nterms$. 
Using these and \reff{composepropb}, we compute the 
power series of $\hat{F_j}$ to 
order $\nterms$. Let $1/R_j$ be the radius of convergence for the
power series of $\hat{F_j}$. 
Now consider evaluating the composition in equation \reff{blockcomposeb}. 
Let $z$ be the argument to $F_j$. 
If $z$ is large compared to $R_j$, then $F_j(z)$ is well approximated 
using the power series of $\hat{F_j}$. 
We introduce a parameter $L>1$ and 
use the power series of $\hat{F_j}$ to compute $F_j(z)$ whenever 
$|z| \ge L R_j$. When $|z| < L R_j$, we just use \reff{blockdefb} to 
compute $F_j(z)$. The argument of $F_j$ is random, and so whether 
or not we can approximate a particular $F_j$ using these power 
series is random.
As part of the algorithm we must compute $R_j$. This is easy.
$R_j$ is the smallest positive number such that $F_j(R_j)$ and 
$F_j(-R_j)$ are both real. 

In addition to the choice of simple curves we use (tilted slits, 
vertical slits, ....), there are three parameters in our algorithm.
$b$ is the number of functions composed in a block.
$n$ is the order at which we truncate our series approximation.
$L$ is the scale that determines when we use series for $F_j$.
The parameter $b$ has little effect on the accuracy of the algorithm 
and we should choose it to make the algorithm run as quickly as possible. 
Eq. \reff{blockcomposeb} suggests that $b$ should vary with $N$ as
$\sqrt{N}$ and experiments bear this out. 

The choice of $n$ involves  a tradeoff of speed vs. accuracy. 
Larger $n$ means more terms in the series, hence slower but more 
accurate computations. 
We typically use $n=12$. 

The parameter $L$ will determine how fast the series converges. 
Roughly speaking, the series will converges at least as 
fast as the geometric series $\sum_n L^{-n}$ .
The choice of $L$ also involves  a tradeoff of speed vs. accuracy. 
Larger $L$ means the series converges faster and so is more accurate. 
But it also means that we use the block functions $F_j$ less 
frequently, and so the computation is slower.
We typically use $L=4$.

\begin{figure}[htb]
\includegraphics{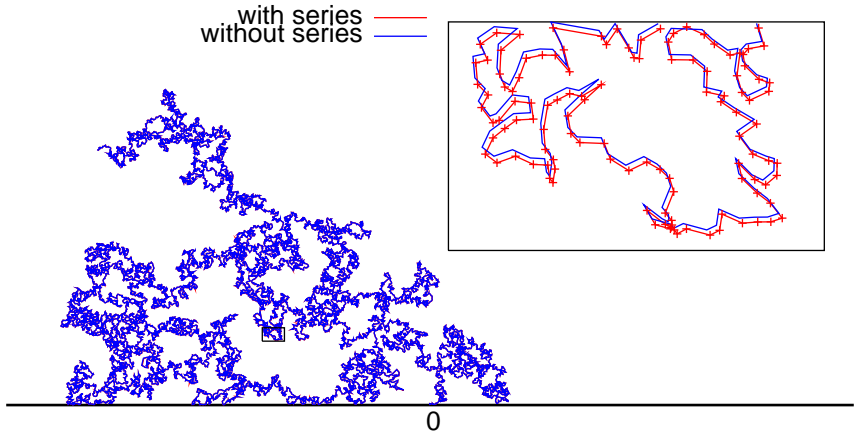}
\caption{Two curves for SLE with $\kappa=6$ are shown. They use the same
Brownian motion sample path but one uses the series approximation and 
the other does not.} 
\label{laurent_error}
\end{figure}

A detailed study of the effects of the choices of $b,n$ and $L$ can 
be found in \cite{tk_sle}. This paper also studies the time to compute a point 
on the curve and finds it is $O(N^p)$ with $p$ approximately $0.4$. 
To illustrate the accuracy of our series approximation we compute an SLE 
curve for $\kappa=6$ with and without the series approximation. 
We use the same Brownian motion sample path for both curves. 
We typically take $n=12$ and $L=4$. With these choices 
the difference between the curves obtained with and without the 
series approximation is extremely small and cannot be seen in plots of 
the curves. If we reduce $n$ to only $6$ we can begin to see the effect 
of the series approximation. 
Figure \ref{laurent_error} shows the two curves we get for $\kappa=6$ 
and the same sample of the driving process when we use $n=6$. 
One can only distinguish the 
difference in the enlargement and even then it is small.

We now consider the algorithm for computing the driving 
function of a given curve. 
The number of operations needed to compute a single $w_{k+1}$ is
proportional to $k$. So to compute all the points $w_{k+1}$, and hence the 
approximation to the driving function, requires a time $O(N^2)$. 
The idea for improving this is the same
as before - we group the functions we are composing into blocks
and approximate the composition $F$ of the functions in a block
using the power series of $\hat{F}$. The only minor difference is that the 
order of the conformal maps in \reff{compose} is the opposite of that in 
\reff{dsle}.
We continue to denote the number of functions in a block by $\blength$. 
Let
\be
H_j = 
h_{jb} \compose h_{jb-1} \compose \cdots \compose h_{(j-1)b+2} \compose 
h_{(j-1)b+1} 
\label{blockdef}
\ee
If we write $k$ as $k=mb+r$ with $0 \le r < b$, then \reff{compose}
becomes
\be
w_{k+1} = 
h_{mb+r}  \compose h_{mb+r-1} \compose \cdots \compose h_{mb+1} \compose 
H_m \compose H_{m-1} \compose \cdots \compose H_1 (z_{k+1})
\label{blockcompose}
\ee
As before, the $h_i$ are relatively simple, but the composition $H_j$ 
cannot be explicitly computed. 
We approximate $h_i$ by the power series of $\hat{h_i}$ 
and compute the approximations to the compositions 
in \reff{blockdef} just once rather than every time we compute a $w_k$. 

Recall that $h_i$ is normalized so that $h_i(\infty)=\infty$ and 
$h_i^\prime(\infty)=1$. It maps $\half$ minus a simple curve near the 
origin to $\half$, sending the tip of the curve to the origin. 
Let $h$ denote such a conformal map. 
Let $R$ be the largest distance from the origin to a point on the curve.
Then $h$ is analytic on $\{z \in \half: |z|>R\}$. 
Since $h$ is real valued on the real axis, the 
Schwarz reflection principle says it may be analytically continued 
to $\{z \in \complex: |z|>R\}$. Moreover, it does not vanish on this domain. 
We could approximate $h$ by its Laurent series about $\infty$, but 
as with the first algorithm it is better to use the power series of  
$\hat{h}(z)=1/h(1/z)$.
Note that the radius of convergence of this power series is $1/R$.

As before, the advantage of working with the power series of $\hat{h}$
is its behavior with respect to composition: 
$ (h_1 \compose h_2) \, \hat{} \, = \hat{h_1} \compose \hat{h_2}$
Our approximation for $h_i(z)$ is to replace $\hat{h_i}(z)$ 
by the truncation of its power series at order $\nterms$ as we did 
in eq. \reff{hat_approx}.
The approximation of $h_i$ and of $H_j$ proceeds as for the first 
algorithm. 
For each $h_i$ we compute the power series of $\hat{h_i}$ to order $\nterms$. 
We then use them to compute the 
power series of $\hat{H}_j$ to order $\nterms$.
As before we introduce a parameter $L>0$. 
Let $1/R_j$ be the radius of convergence for the power series of $\hat{H}_j$. 
Now consider equation \reff{blockcompose}. 
If the argument $z$ of $H_j$ satisfies $|z| \ge L R_j$,
then we approximate $H_j(z)$ using the power series of $\hat{H}_j$.
Otherwise we just use \reff{blockdef} to compute $H_j(z)$. 
The argument of $H_j$ is random, so as before whether 
or not we can approximate a particular $H_j$ by its 
series is random.

We need to compute $R_j$. Consider 
the images of $z_{(j-1)b},z_{(j-1)b+1}, \cdots z_{jb-1}$ 
under the map $H_{j-1} \compose H_{j-2} \compose \cdots \compose H_1$. 
The domain of the conformal map $H_j$ 
is the half-plane $\half$ minus some curve $\Gamma_j$ which passes through
the images of these points. The radius $R_j$ 
is the maximal distance from the origin to a point on $\Gamma_j$. 
This distance should be very close to the maximum
distance from the origin to images of 
$z_{(j-1)b},z_{(j-1)b+1}, \cdots z_{jb-1}$ under 
$H_{j-1} \compose H_{j-2} \compose \cdots \compose H_1$. 
So in our algorithm we approximate $R_j$ by the maximum of these distances.

To compute the driving function without using the power series 
we must compute all the points $w_k$. So if we do not use the 
power series, the time needed is $O(N^2)$. 
The improvement in the speed of the algorithm from using the 
power series approximation is studied in \cite{tk_driving}.
Numerical experiments indicate it is $O(N^p)$ with $p$ approximately 
equal to $1.35$. 

\section{Conclusions and open problems}
\label{sect_conclusions}

We have reviewed numerical methods for taking a driving function and finding the
curve produced by the Loewner equation and for taking a curve in the 
half plane and finding the corresponding driving function. 
Both methods are based on approximating the driving function over short
time intervals by a function for which the Loewner equation may be 
solved explicitly. The solution of the Loewner equation over the entire
interval is then given by a composition of such maps. 
Our numerical studies used as the simple maps the conformal maps 
that produce a vertical slit or a tilted slit in the half plane. 
The difference in the results when we use 
vertical slits or tilted slits is small.
The vertical slit map is considerably faster and simpler
to implement, so we see no reason to use the tilted slit map. 
To simulate SLE effectively it is imperative that the choice 
of time intervals be done in a way that depends on the sample of 
the driving function so that sections of the curve that correspond 
to small changes in capacity are computed accurately.

The speed of both algorithm can be greatly increased by 
using power series approximations of certain analytic functions. 
This approximation is quite accurate and the errors from it are 
insignificant compared to the 
effect of changing the number of points used on the curve 
or compared to the difference between using vertical slits
or tilted slits in the algorithm. 

We end with a discussion of a variety of open problems related to 
these two algorithms. 

We have only discussed the simulation of chordal SLE.
In chordal SLE the random curve goes between two boundary points, 
e.g., the origin and infinity in the half plane. 
In radial SLE the random curve goes between a boundary point 
and an interior point, e.g., the point 1 and the origin in the unit disc. 
The simulation of radial SLE is similar. Can one use the ideas we 
used to speed up the simulation of chordal SLE to speed up the
simulation of radial SLE?

Instead of taking the scaling limit at the critical point, one can 
consider off critical models and take the scaling limit in such a 
way that it has a finite correlation length. What can you say 
about the driving process for this scaling limit ?
For percolation it is know to be rather nasty \cite{nw}. 
See also \cite{bb_off,cfn}.

There are several methods for numerically computing the 
conformal map of a given simply connected domain onto a standard 
domain, like the unit disc. 
One of these methods, the zipper algorithm \cite{kuh,mr}, 
reduces the problem to that
of finding the conformal map from the half plane minus a curve to the 
half plane. So the power series approximation that we use also provides
a faster version of this algorithm. How does this faster version 
compare to other methods for finding the conformal map from a 
simply connected domain to a standard domain \cite{dt,tsai}? 

As discussed in section \ref{sect_sle_intro}, 
it is natural to conjecture that the discrete SLE curve $\hat{\gamma}$ 
introduced in that section converges to the SLE curve for 
$\kappa \le 4$ and converges to the SLE trace which generates the 
SLE hull for $\kappa > 4$. Prove this. Part of the problem 
is figure out the sense in which they converge. 

For the inverse problem of finding the driving function for a given 
curve, there is an analogous convergence question. Show that as the 
number of points used to approximate the curve goes to infinity, the 
computed driving function converges to the true driving function. 
Marshall and Rohde have proved convergence for a particular variant 
of the zipper algorithm \cite{mr}.

As discussed in section \ref{sect_drive_to_curve}, there is a certain 
stability to our approximation of the curve generated by a given 
driving function. Explain this stability. 

\bigskip
\bigskip

\noindent {\bf Acknowledgments:}\\
I thank Don Marshall and Stephen Rohde for  
useful discussions.
Talks and interactions during visits to the 
Banff International Research Station
in March and May of 2005 and to 
the Kavli Institute for Theoretical Physics in September, 2006 
contributed to the research included in these notes.
The opportunity to present this material at the 
2008 Enrage summer school at IHP is warmly acknowledged.
This research was supported in part by the National Science Foundation 
under grants DMS-0201566 and DMS-0501168.

\end{document}